\documentclass[runningheads]{llncs}
\usepackage{graphicx}
\usepackage{pgfplotstable}
\usepackage{listings}
\begin{document}
\title{Enhancing Genetic Improvement Mutations Using Large Language Models}
%
%









\author{Alexander E.I. Brownlee\inst{1}\orcidID{0000-0003-2892-5059} \and
James Callan\inst{2}\orcidID{0000-0002-5692-6203} \and
Karine Even-Mendoza\inst{3}\orcidID{0000-0002-3099-1189}
 \and
Alina Geiger\inst{4}\orcidID{0009-0002-3413-283X}
 \and
Carol Hanna \inst{2}\orcidID{0009-0009-7386-1622}
 \and
Justyna Petke\inst{2}\orcidID{0000-0002-7833-6044}
 \and
Federica Sarro\inst{2}\orcidID{0000-0002-9146-442X}
 \and \\
Dominik Sobania\inst{4}\orcidID{0000-0001-8873-7143}
}
\authorrunning{A. Brownlee et al.}
%
\institute{University of Stirling, UK \and
University College London, UK \and King's College London, UK \and
Johannes Gutenberg University Mainz, Germany}
\maketitle              
\begin{abstract}

Large language models (LLMs) have been successfully applied to software engineering tasks, including program repair. However, their application in search-based techniques such as Genetic Improvement (GI) is still largely unexplored. In this paper, we evaluate the use of LLMs as mutation operators for GI to improve the search process. We expand the \textit{Gin} Java GI toolkit to call OpenAI's API to generate edits for the JCodec tool.
We randomly sample the space of edits using 5 different edit types.
We find that the number of patches passing unit tests is up to $75\%$ higher with LLM-based edits than with standard Insert edits. 
Further, we observe that the patches found with LLMs are generally less diverse compared to standard edits.  
We ran GI with local search to find runtime improvements.
Although many improving patches are found by LLM-enhanced GI, the best improving patch was found by standard GI.

\end{abstract}
\section{Introduction}
As software systems grow larger and more complex, significant manual effort is required to maintain them~\cite{Bohme2017}.
To reduce developer effort in software maintenance and optimization tasks, automated paradigms are essential.
Genetic Improvement (GI)~\cite{Petke2018} applies search-based techniques to improve non-functional properties of existing software such as execution time as well as functional properties like repairing bugs. 
Although GI has had success in industry~\cite{Kirbas2020,sapfix}, it remains limited by the set of mutation operators it employs in the search \cite{petke2023program}.

Large Language Models (LLMs) have a wide range of applications as they are able to process textual queries without additional training for the particular task at hand.
LLMs have been pre-trained on millions of code repositories 
spanning many different programming languages \cite{LLMcode2021}.
Their use for software engineering tasks has had great success~\cite{Hou2023,fan2023large}, showing promise also for program repair~\cite{Sobania2023,Xia2023}. 
Kang and Yoo~\cite{Kang2023} have suggested that there is untapped potential in using LLMs to enhance GI.
GI uses the same mutation operators for different optimization tasks.
These operators are hand-crafted prior to starting the search and thus result in a limited search space.
We hypothesize that augmenting LLM patch suggestions as an additional mutation operator will enrich the search space and result in more successful variants.

In this paper, we conduct several experiments to explore whether using LLMs as a mutation operator in GI can improve the efficiency and efficacy of the search.
Our results show that the LLM-generated patches have compilation rates of $51.32\%$ and $53.54\%$ for random search and local search, respectively (with the \textsc{Medium} prompt category). Previously LLMs (using an LLM model as-is) were shown to produce code that compiled roughly $40\%$ of the time \cite{siddiq2023exploring,xia2023universal}.
We find that randomly sampled LLM-based edits compiled and passed unit tests more often compared to standard GI edits. We observe that the number of patches passing unit tests is up to $75\%$ higher for LLM-based edits than GI \textsc{Insert} edits. However, we observe that patches found with LLMs are less diverse. For local search, the best improvement is achieved using standard GI \textsc{Statement} edits, followed by LLM-based edits. These findings demonstrate the potential of LLMs as mutation operators and highlight the need for further research in this area. 

\section{Experimental Setup}\label{sec:method}
To analyze the use of LLMs as a mutation operator in GI, we used the GPT 3.5 Turbo  model by OpenAI and the GI toolbox Gin~\cite{brownlee2019gin}.
We experimented with two types of search implemented within Gin: random search and local search.
Requests to the LLM using the OpenAI API were via the Langchain4J library, with a temperature of $0.7$.
The target project for improvement in our experiments was the popular JCodec~\cite{jcodec} project which is written in Java. `Hot' methods to be targeted by the edits were identified using Gin's profiler tool by repeating the profiling 20 times and taking the union of the resulting set.

For the random sampling experiments, we set up the runs with statement-level edits (copy/delete/replace/swap from \cite{petke2023program} and insert break/continue/return from \cite{Brownlee2020Injecting}) and LLM edits, generating $1000$ of each type at random. A timeout of $10000$ milliseconds was used for each unit test to catch infinite loops introduced by edits; exceeding the timeout counts as a test failure.
For local search, experiments were set up similarly. There were 10 repeat runs (one for each of the top $10$ hot methods) but the runs were limited to $100$ evaluations resulting in $1000$ evaluations in total, matching the random search. In practice this was $99$ edits per run as the first was used to time the original unpatched code.

\begin{figure}
\lstset{
  frame=lines,
  basicstyle=\small,
  moredelim=[is][\textbf]{_}{_}
}
\begin{lstlisting}
Give me 5 different Java implementations of this method body:
```
<code>
```
This code belongs to project <projectname>.
Wrap all code in curly braces, if it is not already.
Do not include any method or class declarations.
label all code as java.
\end{lstlisting}
\caption{The \textsc{medium} prompt for LLM requests, with line breaks added for readability.}
\label{fig:prompt}
\end{figure}

We experimented with three different prompts for sending requests to the LLM for both types of search: a \textsc{simple} prompt, a \textsc{medium} prompt, and a \textsc{detailed} prompt.
With all three prompts, our implementation requests five different variations of the code at hand.
The \textsc{simple} prompt only requests the code without any additional information.
The \textsc{medium} prompt provides more information about the code provided and the requirements, as shown in Figure \ref{fig:prompt}.
Specifically, we provide the LLM with the programming language used, the project that the code belongs to, as well as formatting instructions.
The \textsc{detailed} prompt extends the \textsc{medium} prompt with an example of a useful change.
This example was taken from results obtained by Brownlee~et~al.~\cite{Brownlee2020Injecting}. 
The patch is a successful instance of the \textsc{insert} edit applied to the jCodec project (i.e., an edit that compiled, passed the unit tests and offered a speedup over the original code).
We use the same example for all the \textsc{detailed} prompt requests used in our experiments; this is because LLMs are capable of inductive reasoning where the user presents specific information, and the LLM can use that input to generate more general statements, further improved in GPT-4 \cite{HAN2023101155}.

LLM edits are applied by selecting a block statement at random in a target `hot' method. This block's content is \verb|<code>| in the prompt. The first code block in the LLM response is identified. Gin uses JavaParser (\url{https://javaparser.org}) internally to represent target source files, so we attempt to parse the LLM suggestion with JavaParser, and replace the original block with the LLM suggestion.


\section{Results}

The first experiment compares standard GI mutations, namely \textsc{Insert} and \textsc{Statement} edits, with LLM edits using differently detailed prompts (\textsc{Simple}, \textsc{Medium}, and \textsc{Detailed}) using Random Sampling. 
Table~\ref{tab:resultsTableNoNoOps} shows results for
all patches as well as for unique patches only. We report how many patches were successfully parsed by JavaParser (named as Valid), how many compiled, and how many passed all unit tests (named as Passed). We excluded patches syntactically equivalent to the original software. Best results are in \textbf{bold}.


\begin{table}[t]
\centering
\caption{Results of our Random Sampling experiment. We exclude patches syntactically equivalent to the original software in this table.
For all and unique patches we report: how many patches passed JavaParser, compiled, and passed all unit tests. 
}
\label{tab:resultsTableNoNoOps}
\resizebox{0.75\linewidth}{!}{
\begin{filecontents*}{data/processedWithoutNoOps.csv}
EditCategory,UniqueValid,UniqueCompiled,UniquePassed,UniquePatches,Valid,Compiled,Passed,Patches
STATEMENT,819,199,80,896,869,227,91,967
INSERT,785,284,161,785,860,295,166,860
LLM SIMPLE,0,0,0,193,0,0,0,1000
LLM MEDIUM,260,183,154,324,463,331,292,645
LLM DETAILED,268,126,110,332,456,250,230,606
\end{filecontents*}
\pgfplotstabletypeset[
    col sep=comma,
    column type=r,
    string type,
    columns={EditCategory, UniquePatches, UniqueValid, UniqueCompiled, UniquePassed, Patches, Valid, Compiled, Passed},
    columns/UniqueValid/.style={column name=Valid},
    columns/UniqueCompiled/.style={column name=Compiled},
    columns/UniquePassed/.style={column name=Passed},
    columns/UniquePatches/.style={column name=Patches},
    every head row/.style={
    before row={
    \hline
    & \multicolumn{4}{c|}{Unique} & \multicolumn{4}{c|}{All}\\
    },
    after row=\hline,
    },	
    every row 0 column 2/.style={
        postproc cell content/.style={
          @cell content/.add={$\bf}{$}
        }
    },
    every row 1 column 3/.style={
        postproc cell content/.style={
          @cell content/.add={$\bf}{$}
        }
    },
    every row 1 column 4/.style={
        postproc cell content/.style={
          @cell content/.add={$\bf}{$}
        }
    },
    every row 0 column 6/.style={
        postproc cell content/.style={
          @cell content/.add={$\bf}{$}
        }
    },
    every row 3 column 7/.style={
        postproc cell content/.style={
          @cell content/.add={$\bf}{$}
        }
    },
    every row 3 column 8/.style={
        postproc cell content/.style={
          @cell content/.add={$\bf}{$}
        }
    },
    every last row/.style={after row=\hline},	
    every column/.style={column type/.add={}{|}},
    every first column/.style={column type = l, column type/.add={|}{|}},	
every row 0 column 0/.style={postproc cell content/.style={@cell
            content=\textsc{Statement}}},
every row 1 column 0/.style={postproc cell content/.style={@cell
            content=\textsc{Insert}}},
every row 2 column 0/.style={postproc cell content/.style={@cell
            content=\textsc{Simple}}},
every row 3 column 0/.style={postproc cell content/.style={@cell
            content=\textsc{Medium}}},
every row 4 column 0/.style={postproc cell content/.style={@cell
            content=\textsc{Detailed}}},
    ]
{data/processedWithoutNoOps.csv}
}
\end{table}

We see that although substantially more valid patches were found with the standard \textsc{Insert} and \textsc{Statement} edits, more passing patches could be found by using the LLM-generated edits. In particular, for the \textsc{Medium}, and \textsc{Detailed} prompts $292$ and $230$ patches passed the unit tests, respectively. For the \textsc{Insert} and \textsc{Statement} edits only $166$ and $91$ passed the unit tests, respectively. 
Anecdotally, the hot methods with lowest/highest patch pass rates differed for each operator: understanding this variation will be interesting for future investigation.

It is also notable that LLM patches are less diverse: over 50\% more unique patches were found by standard mutation operators than the LLM using \textsc{Medium}, and \textsc{Detailed} prompts. With the \textsc{Simple} prompt, however, not a single patch passed the unit tests, since the suggested edits often could not be parsed. Thus detailed prompts are necessary to force LLM to generate usable outputs. 

We investigated further the differences between \textsc{Medium} and \textsc{Detailed} prompts to understand the reduction in performance with \textsc{Detailed} (in the unique patches sets) as \textsc{Medium} had a higher number of compiled and passed patches. 
In both prompt levels, the generated response was the same for 42 cases (out of the total unique valid cases). However, \textsc{Detailed} tended to generate longer responses with an average of 363 characters, whereas \textsc{Medium} had an average of 304 characters. 
We manually examined several \textsc{Detailed} prompt responses, in which we identified some including variables from other files, potentially offering a significant expansion of the set of code variants GI can explore.

The second experiment expands our analysis, comparing the performance of the standard and LLM edits with Local Search. 
Table~\ref{tab:resultsLSTableNoNoEdit} shows the results of the Local Search experiment. We report the number of compiling and passing patches as well as the number of patches were runtime improvements were found. Furthermore, we report the median and best improvement in milliseconds (ms). In the table, we excluded all empty patches. As before, best results are in \textbf{bold}. 


\begin{table}[t]
\centering
\caption{Local Search results. We exclude all empty patches. We report how many patches compiled, passed all unit tests, and how many led to improvements in runtime. We report best improvement found and median improvement among improving patches.}
\label{tab:resultsLSTableNoNoEdit}
\resizebox{0.75\linewidth}{!}{
\begin{filecontents*}{data/processedLSWithoutNoEdit.csv}
EditCategory,UniqueCompiled,UniquePassed,UniquePatches,Compiled,Passed,Patches,ImprovFound,BestImprov,Median
STATEMENT,189,104,896,213,105,990,71,508.0,137.0
INSERT,314,176,747,414,264,948,136,313.0,81.0
LLM SIMPLE,2,2,80,2,2,990,2,176.0,137.5
LLM MEDIUM,270,262,617,530,520,990,164,395.0,75.5
LLM DETAILED,158,149,480,379,369,990,196,316.0,95.0
\end{filecontents*}
\pgfplotstabletypeset[
    col sep=comma,
    column type=r,
    string type,
    columns={EditCategory, Patches, Compiled, Passed, ImprovFound, BestImprov, Median},
    columns/ImprovFound/.style={column name=ImpFound},
    columns/BestImprov/.style={column name=BestImp(ms)},
    columns/Median/.style={column name=Median(ms)},
    every head row/.style={
    before row=\hline,
    after row={
    \hline
    },
    },	
    every row 3 column 2/.style={
        postproc cell content/.style={
          @cell content/.add={$\bf}{$}
        }
    },
    every row 3 column 3/.style={
        postproc cell content/.style={
          @cell content/.add={$\bf}{$}
        }
    },
    every row 4 column 4/.style={
        postproc cell content/.style={
          @cell content/.add={$\bf}{$}
        }
    },
    every row 0 column 5/.style={
        postproc cell content/.style={
          @cell content/.add={$\bf}{$}
        }
    },
    every row 2 column 6/.style={
        postproc cell content/.style={
          @cell content/.add={$\bf}{$}
        }
    },
    every last row/.style={after row=\hline},	
    every column/.style={column type/.add={}{|}},
    every first column/.style={column type = l, column type/.add={|}{|}},	
every row 0 column 0/.style={postproc cell content/.style={@cell
            content=\textsc{Statement}}},
every row 1 column 0/.style={postproc cell content/.style={@cell
            content=\textsc{Insert}}},
every row 2 column 0/.style={postproc cell content/.style={@cell
            content=\textsc{Simple}}},
every row 3 column 0/.style={postproc cell content/.style={@cell
            content=\textsc{Medium}}},
every row 4 column 0/.style={postproc cell content/.style={@cell
            content=\textsc{Detailed}}},
    ]
{data/processedLSWithoutNoEdit.csv}
}
\end{table}

Again, we see that more patches passing the unit tests could be found with the LLM using the \textsc{Medium}, and \textsc{Detailed} prompts. In addition, more improvements could be found by using the LLM with these prompts. Specifically, with \textsc{Medium} and \textsc{Detailed}, we found $164$ and $196$ improvements, respectively, while we only found $136$ with \textsc{Insert} and $71$ with \textsc{Statement}. The best improvement could be found with $508$~ms with the \textsc{Statement} edit. The best improvement found using LLMs (using the \textsc{Medium} prompt) was only able to improve the runtime by $395$~ms. 
We also examined a series of edits in Local Search results to gain insights into the distinctions between \textsc{Medium} and \textsc{Detailed} prompts due to the low compilation rate of \textsc{Detailed} prompt's responses. In the example, a sequence of edits aimed to inline a call to function \textsc{clip}. The \textsc{Detailed} prompt tried to incorporate the call almost immediately within a few edits, likely leading to invalid code. On the other hand, the \textsc{Medium} prompt made less radical changes, gradually refining the code. It began by replacing the ternary operator expression with an if-then-else statement and system function calls before eventually attempting to inline the \textsc{clip} function call.



\section{Conclusions and Future Work}
Genetic improvement of software is highly dependent on the mutation operators it utilizes in the search process. To diversify the operators and enrich the search space further, we incorporated a Large Language Model (LLM) as an operator.

\textbf{Limitations.}
To generalise, future work should consider projects besides our target, jCodec.
Our experiments used an API giving us no control over the responses generated by the LLM or any way of modifying or optimizing them. Though we did not observe changes in behaviour during our experiments, OpenAI may change the model at any time, so future work should consider local models.
We experimented with only three prompt types for LLM requests and within this limited number of prompts found a variation in the results.
Finally, our implementation for parsing the responses from the LLMs was relatively simplistic.
However, this would only mean that our reported results are pessimistic and an even larger improvement might be achieved by the LLM-based operator.

\textbf{Summary.}
We found that, although more valid and diverse patches were found with standard edits using Random Sampling, more patches passing the unit tests were found with LLM-based edits. For example, with the LLM edit using the \textsc{Medium} prompt, we found over 75\% more patches passing the unit tests than with the classic \textsc{Insert} edit. In our Local Search experiment, we found the best improvement with the \textsc{Statement} edit (508~ms). The best LLM-based improvement was found with the \textsc{Medium} prompt (395~ms). Thus there is potential in exploring approaches combining both LLM and `classic' GI edits.

Our experiments revealed that the prompts used for LLM requests greatly affect the results. Thus, in future work, we hope to experiment more with prompt engineering. It might also be helpful to mix prompts: e.g., starting with \textsc{medium} then switching to \textsc{detailed} to make larger edits that break out of local minima. Further, the possibility of combining LLM edits with others such as standard copy/delete/replace/swap or PAR templates~\cite{Kim} could be interesting. Finally, we hope to conduct more extensive experimentation on additional test programs.

\vspace{0.8em}

\textbf{Data Availability.}
The code, LLMs prompt and experimental infrastructure, data from the evaluation, 
and results are available as open source at~\cite{TooL_artifact_2023}. The code is also under the `llm' branch of \url{github.com/gintool/gin} (commit 9fe9bdf; branched from master commit 2359f57 pending full integration with Gin).

\vspace{0.8em}

\textbf{Acknowledgements}
UKRI EPSRC EP/P023991/1 and ERC 741278.

\bibliographystyle{splncs04}
\bibliography{references}

\end{document}